\documentclass[12pt]{article}
\pdfoutput=1

\usepackage{epsfig}
\usepackage{graphicx}
\usepackage{epstopdf}
\usepackage{amsfonts}
\usepackage{amssymb}
\usepackage{amsthm}
\usepackage{amsmath}
\usepackage{multirow}
\numberwithin{equation}{section} 
\newcommand{\newc}{\newcommand}
\newc{\ra}{\rightarrow}
\newc{\lra}{\leftrightarrow}
\newc{\be}{\begin{equation}}
\newc{\ee}{\end{equation}}
\newc{\bg}{\begin{gathered}}
\newc{\eg}{\end{gathered}}
\newc{\bs}{\begin{split}}
	\newc{\es}{\end{split}}
\newc{\ba}{\begin{eqnarray}}
\newc{\ea}{\end{eqnarray}}
\newc{\ov}{\overline}
\newc{\pa}{\partial}
\newc{\D}{\Delta}

\newc{\nn}{\nonumber}

\newc{\tref}[1]{Table \ref{#1}}
\newc{\eref}[1]{Equation \eqref{#1}}
\newc{\fref}[1]{Figure \ref{#1}}
\newc{\sref}[1]{Section \ref{#1}}
\newc{\su}[1]{$SU(#1)$}
\newc{\bm}[1]{\mathbf{#1}}

\begin{document}
	\begin{titlepage}
		
		\vspace*{0.7cm}

		\begin{center}
			{	\bf
				Discrete  Flavour Symmetries from the Heisenberg group 
			 }
			\\[12mm]
			E. G. Floratos$^{\it a,b}$~\footnote{E-mail:\texttt{mflorato@phys.uoa.gr}},
			G.K. Leontaris$^{c}$~\footnote{E-mail: \texttt{leonta@uoi.gr}}
			\\[-2mm]
			
		\end{center}
		\vspace*{0.50cm}
			\centerline{$^{a}$ \it
				Theory Division, CERN,}
			\centerline{\it
				1211 Geneva 23, Switzerland }
		\vspace*{0.2cm}
		\centerline{$^{b}$ \it
			 Physics Department, University of Athens,}
		\centerline{\it
		Zografou GR-15784  	 Athens, Greece }
		\vspace*{0.2cm}
		\centerline{$^{c}$ \it
			Physics Department, Theory Division, Ioannina University,}
		\centerline{\it
			GR-45110 Ioannina, Greece}
		\vspace*{1.20cm}
		
		\begin{abstract}
			\noindent
			Non-abelian discrete symmetries are of particular importance in model building.
	    	They are mainly invoked to explain the various fermion mass hierarchies and 
	    	 forbid dangerous superpotential terms. In string models they are usually 
	    	 associated to the geometry of the compactification manifold and more particularly
	    	 to the magnetised branes in toroidal compactifications. Motivated by these
			 facts, in this note we propose a unified framework to construct representations of
			 finite discrete family groups based on the automorphisms of the discrete and finite Heisenberg group. 
			  We focus in particular, in  the $PSL_2(p)$ groups which contain the phenomenologically interesting cases.

		\end{abstract}
		
	\end{titlepage}
	
	\thispagestyle{empty}
	\vfill
	\newpage
	
	\setcounter{page}{1}

\section{Introduction}

Non-abelian discrete symmetries play a prominent r\^ole in model building. Among other objectives, more than a decade ago, 
they have been widely  used to interpret the neutrino data in various extensions of the Standard
 Mode~\cite{Altarelli:2010gt,Ishimori:2010au,King:2013eh}. 
 The ensuing  years there were attempts to construct them in the context of  string theory models.
 Within this framework,  important activity has been focused on elaborating predictions for physically measurable  
 quantities such as mass textures and CP-violating matrices. Indeed, (non-abelian) discrete symmetries  in string 
 theory  emerge in the context of various compactifications  and recently they have attracted considerable 
attention~\cite{BerasaluceGonzalez:2012vb}-\cite{Grimm:2015ona}.
In fact, it has been realised that they can act as family symmetries
 which restrict the arbitrary Yukawa parameters of the 
superpotential and lead to acceptable quark and lepton masses and mixing. 
Moreover,  they can suppress undesired proton
decay operators and various -yet unobserved- flavour violating interactions.

More recently, the implementation of the idea of discrete symmetries has also been 
considered in  F-theory constructions.
In F-theory~\cite{Vafa:1996xn} the elliptically fibred space consists of 
a $K3$ manifold with a torus attached at each point.  The $\tau$-modulus of the
torus is defined in terms of the two scalar fields of the type IIB string theory and the fibration is described
by the Weierstra\ss\, model. According to the standard interpretation, the associated geometric singularities
(classified as ADE types) are linked to the gauge symmetries of the effective models.
The highest singularity of the elliptic fibration is described by the $E_8$ exceptional symmetry, so that 
ordinary successful GUT symmetries such as $SU(5)$ and $SO(10)$ are easily
embedded~\cite{Beasley:2008kw} in the maximal group $E_8$ and correspond to a particular divisor of the internal manifold. 
Hence, the remaining symmetry can in principle accommodate some suitable non-abelian discrete
group which could act as a family symmetry. If, for example, the GUT model is  $SU(5)$ which 
is the minimal symmetry accommodating the Standard Model,  
then the commutant with respect to $E_8$ is also $SU(5)$ (denoted usually as perpendicular, $SU(5)_{\perp}$, to the GUT).
 The latter naturally incorporates phenomenologically viable non-abelian discrete 
 groups~\cite{Karozas:2015zza}, such as $S_n,A_n$ where usually  $n\le 5$ and more generally  $PSL_2(p)$, where   $p\le 11$.

 The last couple of years, in the context of F-theory, several works focused also in the low energy implications of 
 the two torus geometry, in a  different approach.  In general, discrete symmetries in these constructions are of 
 Abelian nature.   Such cases are the Torsion part of the Mordell-Weil group of rational points on elliptic curves  and more 
generally the Tate-Shafarevich group which has been shown to determine the discrete symmetries arising in F-theory~\cite{Braun:2014oya}.

Furthermore, there exist cases in string theory~\cite{BerasaluceGonzalez:2012vb} where non-abelian finite 
groups may emerge as well. In this context a class of non-Abelian 
discrete symmetries may arise from  discrete isometries
of the torus geometry, on  which  the Heiseberg group has a  natural action. 
Discrete non-abelian symmetries are also realised in magnetised D-brane models in toroidal 
compactifications~\cite{BerasaluceGonzalez:2012vb,Abe:2009vi}.
Along these lines, explicit F-theory constructions have also recently appeared~\cite{Grimm:2015ona}.
Let us finally note that finite groups as subgroups of continuous non-abelian symmetries have 
been discussed and classified in an orbifold context~\cite{Hanany:1999sp}.

Motivated by the above facts, in the present work we will develop a unified method for the construction
of the smaller non-trivial representations of certain finite groups. 
Because their main r\^ole in particle physics is to  discriminate the three fermion families 
we are focusing mainly  on the $PSL_2(p)$ groups  possessing triplet representations.

\section{The non-abelian discrete groups  $SL_2(p)$ and $PSL_2(p)$}

Among the various discrete symmetries used to interpret the fermion mass hierarchy
are the special linear groups $SL_2(p)$\cite{Tanaka,Humphreys}  and their corresponding projective restrictions $PSL_2(p)$.
Some of these groups coincide with the symmetries of regular polyedra in three or higher dimensional 
space dimensions. 

Not all the representations of these groups are relevant for model  building. Because 
 only three families of fermions exist in nature, 
only the groups with particular representations related to  Yukawa and gauge couplings 
are considered in the literature~\cite{Altarelli:2010gt,Ishimori:2010au,King:2013eh}.

Most of the phenomenologically viable cases, are included in the  projective linear groups $PSL_2(p)$
 for $p=3,5,7 $ and $11$. All of them support triplets which are suitable representations to
accommodate the three fermion generations. More specifically, 
for  $p=3$ we obtain $PSL_2(3)$ which is isomorphic to the alternating group $A_4$.  
Furthermore, the group $PSL_2(5)$  is isomorphic to the smallest non-abelian simple group $A_5$. 
The case of  $PSL_2(7)$ is also phenomenologically interesting~\cite{Luhn:2007yr,Luhn:2007sy,King:2009ap}.

The $PSL_2(p)$ groups for values $p\le 11$  can be naturally incorporated in an F-theory context. Indeed,
for the most common grand unified (GUT) models such as $SU(5), SO(10), E_6$ 
embedded in the $E_8$ singularity, the possible  gauge groups which could
act as family symmetries can be read off from the embedding formula
\ba
 E_8\supset \frac{E_n\times SU(m)}{Z_m},\;\; n+m=9
\ea
Therefore, we have the following cases
\ba
E_8&\supset&E_6\times SU(3)_{\perp}\label{su3}\\
E_8&\supset&E_5\times SU(4)_{\perp}\cong SO(10)\times SU(4)_{\perp}\label{E5chain}\\
E_8&\supset&SU(5)\times SU(5)_{\perp}\label{su5}
\ea
From the above, we see that the corresponding flavour discrete groups should be embedded in $SU(3)_{\perp}, SU(4)_{\perp}$ and $SU(5)_{\perp}$.
Indeed, $A_{4,5}$ are subgroups of $SO(3)\sim SU(2)$,  $PSL_2(7)$ is a subgroup of $SU(3)$ and $PSL_2(11)$ is contained
 in  $SU(5)$.

In this work, we present a unified approach for constructing the explicit relevant  representations of these 
groups. The reason is that  in the literature, up to our knowledge,  
while only explicit ad-hoc constructions have been presented,
 a systematic use of the theory of the
representation of these particular groups does not exist yet.

\section{Definition of $SL_2(p)$ and $PSL_2(p)$ groups}

The  $SL_2(p)$ group is defined in the simplest way as a group of $2\times 2$ matrices  with elements integers
modulo $p$, where $p$ is a prime integer, and determinant one modulo $p$.
These groups usually are generated by two elements which obey certain conditions and these define what is called a presentation
of the group. These conditions depend on $p$ but there is a universal presentation given by two generators 
which are conjugate one to another under the matrix 
\[    a=\left(\begin{array}{cc}0&-1\\1&0\end{array}\right)\]
These universal  elements  (for any value of $p$) are  the following two matrices
\[    L=\left(\begin{array}{cc}1&0\\1&1\end{array}\right),\;  
R=\left(\begin{array}{cc}1&-1\\0&1\end{array}\right)\] 
We can readily observe that each one of them generate an abelian group of order $p$ and 
that they satisfy  the Braid relation
\[  RLR= LRL \]
In the literature another basis of generators is used which is called Artin's presentation.
These are  defined through the Braid generators, as
 \[a=RLR,\; b=RL \]
and so they are 
\be    
 a=\left(\begin{array}{cc}0&-1\\1&0\end{array}\right),\;  
b=\left(\begin{array}{cc}0&-1\\1&1\end{array}\right)\label{abgen}
\ee 
They satisfy the relations
\be  a^2 = b^3=-I\label{abrel}\ee 
We can invert the relation of the two sets of generators as
\be R=b^{-1}a,\;  L=a^{-1}b^2   \label{RLba} 
\ee
Apart from equations (\ref{abrel}),  the presentation in terms of $a,b$  contains  additional relations 
depending on the value of $p$.

The group  $SL_2(p)$ has a normal subgroup of order two $Z_2={\{1,-1\}}$, so the coset space 
$SL_2(p)/{\{1,-1\}}$ is a group which is called the projective group $ PSL_2(p)$.

Our aim is to construct some basic non-trivial irreducible representations of $PSL_2(p)$  out of which all others -physically
relevant- are generated by tensor products. To this end, we are going to use a particular 
representation of $SL_2(p)$ which is known as the Weil metaplectic representation. In the physics
literature is has been introduced by the work of Balian and Itzykson~\cite{BI}.
Many details of this particular unitary  representation with various applications has been presented in~\cite{Athanasiu:1994fv,Athanasiu:1998cq}.
As it will be shown in the next section, this representation is reducible in  two irreducible unitary
representations of dimensions $\frac{p+ 1}2$ and $\frac{p- 1}2$. Thus, we obtain discrete subgroups of the unitary groups 
$SU(\frac{p\pm 1}2)$.  For example, when $p=3$ we obtain discrete subgroups of $SU(2)$ and $U(1)$, for 
$p=5$ we get discrete subgroups of $SU(3)$ and $SU(2)$ and so on.

\section{Heiseberg-Weyl group ${ HW}_p$ and the metaplectic representation of  $SL_2(p)$ }

The Finite  Heisenberg group  ${ HW}_p$~\cite{JS}, is defined as the set of $3\times 3$ matrices of the form
\be 
 g(r,s,t)= \left(\begin{array}{ccc}1&0&0\\r&1&0\\t&s&1\end{array}\right)
\label{HeisGroup}
\ee
where $r,s,t$ belong to $ \mathbb{Z}_p$ (integers modulo $p$), where the multiplication
of two elements is carried modulo $p$. 

\noindent 
When  $p$ is a prime integer   there is a unique $p$-dimensional  unitary irreducible and faithful representation of this group,
given  by the following matrices
 \ba 
 J_{r,s,t}&=&\omega^t\,P^r Q^s
 \ea
where $\omega= e^{2\pi i/p}$, i.e. the $p^{th}$ primitive  root of unity and the matrices $P,Q$ are defined as
\ba 
P_{kl}&=&\delta_{k-1,l}\\
Q_{kl}&=&\omega^{k}\delta_{kl}
\ea 
where $k,l=0,\dots, p-1$. 

It is to be  observed that, if  $\omega$  is replaced with  $\omega^k$,
for $k=1,2,...,p-1$ all the relations above remain intact.
Since $p$ is prime all the resulting representations are $p$-dimensional and inequivalent.

The matrices  $P,Q$  satisfy the fundamental Heisenberg commutation relation of
Quantum Mechanics in an exponentiated form
\be 
Q\,P =\omega\, P\,Q \label{QPC}
\ee 
In the above,  $Q$ represents the  position  operator on the circle  $\mathbb{Z}_p$  of the $p$ roots of unity
and $P$ the corresponding momentum operator. These two operators are related by the
diagonalising unitary matrix  $F$ of $P$,
\be  
QF =FP \label{QPF}
\ee 
so $F$ is the celebrated Discrete Fourier Transform  matrix 
\be 
  F_{kl}= \frac{1}{\sqrt{p}} \omega^{kl},\;{\rm  with }  \; k,l=0,\dots, p-1\label{FT}
  \ee

An important subset of ${ HW}_p$ consists of  the magnetic translations 
\be 
J_{r,s}= \omega^{rs/2}P^rQ^s
\ee 
with $r,s=0,\dots, p-1$.
These matrices are unitary ($J_{r,s}^{\dagger}=J_{-r,-s}$) and traceless, and  they form
a basis for the Lie algebra of $SL(p, \mathbb{C})$. They satisfy the important relation
\be 
J_{r,s}J_{r',s'} = \omega^{(r's-rs')/2}J_{r+r',s+s'}\label{JrsC}
\ee 
This relation implies that the magnetic translations form a projective representation
of the translation group  $ \mathbb{Z}_p\times   \mathbb{Z}_p$. The factor of $1/2$ in the exponent
of (\ref{JrsC}) must be taken modulo $p$.

The $SL_2(p)$ appears here as the automorphism group of magnetic translations and this defines the Weil's 
metaplectic representation. If we consider the action of an element ${\tiny A= \left(\begin{array}{cc}a&b\\c&d\end{array}\right)}$
on the coordinates $(r,s)$ of the periodic torus  $ \mathbb{Z}_p\times   \mathbb{Z}_p$, this induces a unitary automorphism $U(A)$ on the
magnetic translations, since the representation of Heisenberg group is unitary and irreducible,
\[   U(A)J_{r,s}U^{\dagger}(A)= J_{r',s'}  \]
where $(r',s')$ are given by
\ba 
(r',s') &=& (r,s) \left(\begin{array}{cc}a&b\\c&d\end{array}\right)\label{JrsC1}
\ea 
This relation determines $U(A)$ up to a phase and in the case of $A\in SL_2(p)$,  the phase can be
fixed to give  an exact (and not projective) unitary representation of $ SL_2(p)$.

\noindent
The detailed formula of $U(A)$ has been given by Balian and Itzykson~\cite{BI}. Depending on the specific 
values of the $a,b,c,d$ parameters of the matrix $A$, we distinguish the following cases:
\ba  
\delta\ne 0: && U(A)=\frac{\sigma(1)\sigma(\delta)}{p}\sum_{r,s}\omega^{[br^2+(d-a)rs-cs^2]/(2\delta)}J_{r,s}
\label{s1}\\
\delta = 0,\; b\ne 0:&& U(A)=\frac{\sigma(-2b)}{\sqrt{p}}\sum_{s}\omega^{s^2/(2b)}J_{s(a-1)/b,s}
\label{s2}
\\
\delta = b=0,\; c\ne 0:&& U(A)=\frac{\sigma(2c)}{\sqrt{p}}\sum_{r}\omega^{-r^2/(2c)}P^r
\label{s3}\\
\delta = b=0= c= 0:&&U(1)=I
\ea
where $\delta =2-a-d$ and $\sigma(a)$
is
the Quadratic Gauss sum given by
\be 
\sigma(a) =\frac{1}{\sqrt{p}}\,\sum_{k=0}^{p-1}\omega^{ak^2} =(a|p)\times \left\{\begin{array}{cc}1&
{\rm for}\, p=4k+1\\i&{\rm for}\, p=4k-1\end{array}\right.
\ee 
while the Legendre symbol takes the values $(a|p)=\pm 1$ depending whether $a$ is or is  not a square modulo $p$.

It is possible to perform explicitly the above Gaussian sums noticing that 
\be   
(J_{r,s})_{k,l} = \delta_{r,k-l} \omega^{\frac{k+l}2s}   \label{Jrs}
\ee 
where all indices take the values $k,l,r,s=0,\dots, p-1$. 
This has been done in ~\cite{Athanasiu:1994fv,Athanasiu:1998cq}. In the case 
$\delta =2-a-d\ne 0$ and $c\ne 0$, the result is
\ba  
\delta \ne 0: && U(A)_{k,l}=\frac{(-2c|p)}{\sqrt{p}}\,\times \left\{\begin{array}{c}1\\-i\end{array}\right\}
\omega^{-\frac{ak^2-2kl+dl^2}{2c}}
\label{UAelms}
\ea

If $c=0$, then we transform the matrix $A$ to one with $c\ne 0$. The other cases $\delta =0$ can be worked out easily 
using the matrix elements of $J_{r,s}$ given in~(\ref{Jrs}).

It is interesting to notice that redefining $\omega$ to become $\omega^k$ for $k=1,2,...,p-1$, the matrix $U(A)$ 
transforms to the matrix $U(A_k)$, where $A_k$  is the $2\times 2$ matrix
 ${\tiny A_k= \left(\begin{array}{cc}a&b k\\c/k&d\end{array}\right)}$, which 
belongs to the same conjugacy class with  $A$ as long as $k$ is  a quadratic residue.
If $k=p-1$  we pass from the representation  $U(A)$  to the complex conjugate one  $U(A)^*$.

The Weyl representation presented above, provides the interesting result that the unitary matrix corresponding to the
$SL_2(p)$  element ${\tiny a=\left(\begin{array}{cc}0&-1\\1&0\end{array}\right)}$  is -up to a phase- the 
Discrete Finite Fourier Transform (\ref{FT})
\[ U(a) =(-1)^{k+1}  i^n F    \]
where $n=0$ for $p=4k+1$ and $n=1$ for $p=4k-1$. 

The Fourier Transform matrix generates a fourth order abelian group with elements
\be 
F,\; F^2=S, \;F^3=F^*,\; F^4=I\label{Fgroup}
\ee

The matrix $S$ represents the element ${\tiny a^2=\left(\begin{array}{cc}-1&0\\0&-1\end{array}\right)}$ .
Its matrix elements are 
\ba 
S_{k,l} &=&\delta_{k,-l},\;\; k,l=0,\dots p-1\label{Sij}\\
U(a^2)_{k,l}&=&i^{2n}S_{k,l} =(-)^n\delta_{k,-l},\;\; k,l=0,\dots p-1
\label{UandS}
\ea
Because the action of $S$ on $J_{r,s}$ changes the signs of $r,s$, while  $\forall A\in SL_2(p)$ 
the unitary matrix $U(A)$  depends quadratically on $r,s$ in the sum (\ref{s1}), it turns out 
that $S$ commutes with all $U(A)$. Moreover, $S^2=I$ and we can construct two projectors 
\[ P_+=\frac 12( {I+S}),\; P_- = \frac 12 (I-S) \]
with dimensions of their invariant subspaces $\frac{p+1}2$ and $\frac{p-1}2$ correspondingly. So the Weil 
$p$-dimensional representation is the direct sum of two irreducible  
unitary representations 
\[ U_+(A)= U(A) P_+, \;  U_-(A)=U(A)P_- \]

To obtain the block diagonal form of the above matrices $U_{\pm}(A)$, we rotate with the orthogonal 
matrix of the eigenvectors of $S$.
This $p$-dimensional orthogonal matrix, dubbed here $O_p$,
 can be obtained in a maximally symmetric form (along the diagonal as well as along the anti-diagonal) 
 using the eigenvectors of $S$ in the following order:
In the first $(p+1)/2$ columns we put the eigenvectors of $S$ of eigenvalue equal to 1,  and in the next
$(p-1)/2$ columns the eigenvectors of eigenvalue equal to $-1$ in the specific order given below:
\ba 
({e_0})_k&=& \delta_{k0},\\
({e^+_j})_k&=&\frac{1}{\sqrt{2}}(\delta_{k,j}+\delta_{k,-j}),\; j=1,\dots, \frac{p-1}2\\
({e^-_j})_k&=&\frac{1}{\sqrt{2}}(\delta_{k,j}-\delta_{k,-j}),\; j=\frac{p+1}2,\dots, p
\ea 
where $k=0,\dots, p-1$.

Different orderings of eigenvectors may lead to different  forms of the matrices $U_{\pm}(A)$.
The so obtained orthogonal matrix $O_p$ has the property 
\[  O_p^2 =I \]
due to its symmetric form.

The final block diagonal form of $U_{\pm}(A)$ is obtained through an $O_p$ rotation
\be   V_{\pm}(A)= O_p U(A)_{\pm} O_p \label{VM}  \ee

\section{The construction of the $SL_2(p)$ generators}

In this section we are going to give explicit expressions for the $SL_2(p)$ generators  
 $a,b$  for any value of $p$ in the $\frac{p\pm 1}2$ irreducible representations. 
We will also consider the corresponding matrix expressions of the projective group $PSL_2(p)$.

According to the above construction the two generators $a,b$ have the following unitary matrix representations
\ba 
 U_{\pm} (a) &=& (-1)^{k+1} i^nF\frac{I\pm S}2 = (-1)^{k+1} i^n\frac 12(F\pm F^*)  
 \ea 
with matrix elements
\[ \left[  U_{\pm} (a)\right]_{k,l} =(-1)^{k+1}i^n\frac{1}{2\sqrt{p}} \left(\omega^{kl}\pm \omega^{-kl}\right)     \]
where, as noted previously, $n=0$ for $p=4k+1$ and $n=1$ for $p=4k-1$.
 
In order to bring this in the block diagonal form (\ref{VM}) we need to perform a rotation with $O_p$:
\be 
{ A}^{[\frac{p\pm 1}2]}= O_p U_{\pm}(a) O_p\,.
\ee

For the second generator, $b$, given in (\ref{abgen}) we obtain 
\ba 
U(b)_{k,l} &=&\frac{1}{\sqrt{p}} (-1)^{\frac{p^2-1}8}\left\{\begin{array}{c}1\\i\end{array}  \right\} \omega^{-\frac{l^2}2+kl}
\ea
and so,
\[{U_{\pm}}(b)_{k,l} = \frac{1}{2\sqrt{p}} (-1)^{\frac{p^2-1}8}\left\{\begin{array}{c}1\\i\end{array}  \right\}
 \left( \omega^{-\frac{k^2}2+kl}\pm \omega^{-\frac{l^2}2-kl} \right)    \]
where  $k,l=0,\dots , p-1$.
As previously, in order to get the block diagonal form we have to rotate the so obtained matrix with $O_p$:
\be 
{B}^{[\frac{p\pm 1}2]}= O_p U_{\pm}(b) O_p\,.\label{bb}
\ee

Our final goal is to obtain some basic representations of $PSL_2(p)$ which will be used to build higher dimensional
ones. We observe that the difference between $SL_2(p)$ and $PSL_2(p)$ in the defining relations of
generators $a$ and $b$ is that, for $SL_2(p)$ one has to take $a^2=b^3=-I$  while for $PSL_2(p)$ we have the
relations $a^2=b^3=I$. This last requirement comes from the different action of $SL_2(p)$ and  $PSL_2(p)$
which are linear and M\"obius correspondingly.

We can obtain irreducible  representations of $PSL_2(p)$ from 
irreducible representations of $SL_2(p)$ in the following way.
Taking into consideration the above observation we must find
the representations of $SL_2(p)$ for which $({ A}^{[\frac{p\pm 1}2]})^2=({B}^{[\frac{p\pm 1}2]})^3=I$.
We can easily check  that this happens  for the 
$\frac{p+1}{2}$ dimensional representation  only  when $p=4 k+1$, and for the $\frac{p-1}{2}$ one  only when
$p=4 k-1$. This way, we get $(2k+1)$ and $(2k-1)$-dimensional  irreps of  $PSL_2(p)$ correspondingly.

\section{Examples: The cases $p=3,5,7$ }

In this section using  the  method described above, we will present
examples, considering the cases $p=3,5$ and $p=7$. It is straightforward
to construct similar representations of higher values of $p$.

\subsection{ The case $p=3$ }

The resulting group is $SL_2(3)$ which  has 24 elements,  while its  projective subgroup $PSL_2(3)$ 
has 12 elements and is isomorphic to $A_4$, the symmetry group of the even permutations of four objects,\footnote{$A_4$ 
is suitable to reproduce the TriBi-maximal mixing to leading order in the neutrino sector and has been discussed 
in many works  including~\cite{Ma:2001dn}. See reviews [1,2,3] for a complete list of related papers.} 
or the symmetry group of the Tetrahedron $T$. The symmetry groups of the Cube and the Octahedron is $S_4$ which 
is isomorphic to $PGL_2(3)$, the automorphism group of $SL_2(3)$.

\noindent
The generators in the doublet representation are the following: 
the  $A^{[2]}$-representation is
\ba 
A^{[2]}&=&-\frac{i}{\sqrt{3}}
\left(\begin{array}{cc}
	1&\sqrt{2}\\
\sqrt{2}&\eta+\eta^2\\
\end{array}\right)=-\frac{i}{\sqrt{3}}
\left(\begin{array}{cc}
	1&\sqrt{2}\\
\sqrt{2}&-1\\
\end{array}\right)
\ea 
where we have used the fact that $1+\eta+\eta^2=0$.
The  representation  $B^{[2]}$ is given by
\ba 
B^{[2]}&=&-\frac{i}{\sqrt{3}}
\left(\begin{array}{cc}
	1&\sqrt{2}\eta\\
	\sqrt{2}&1+\eta^2\\
\end{array}\right) =-\frac{i}{\sqrt{3}}
\left(\begin{array}{cc}
	1&\sqrt{2}\eta\\
	\sqrt{2}&-\eta\\
\end{array}\right)\,.
\ea 
They satisfy the $SL_2(3)$ relations 
\be 
{A^{[2]}}^2={B^{[2]}}^3=({A^{[2]}B^{[2]}})^3=-I
\ee

\noindent
The singlet representations are
\ba
A^{[1]}&=&-\frac{i}{\sqrt{3}} (\eta-\eta^2)
\label{S3A}
\\ 
{B^{[1]}}&=&-\frac{i}{\sqrt{3}} (\eta^2-1)
\label{S3B}
\ea 

\noindent
The defining relations are satisfied
\be 
{A^{[1]}}^2={B^{[1]}}^3=({A^{[1]}B^{[1]}})^3=1
\ee 
and are consistent with the   $PSL_2(3)\sim A_4$.
From (\ref{S3A},\ref{S3B}) we deduce
\[  A^{[1]}\cdot  B^{[1]} \equiv \eta,\; \; (A^{[1]}\cdot  B^{[1]})^2\equiv \eta^2 \]
so that we can define the singlet representations with the 
standard multiplications rules:
\ba 
1' &:& s_{1'}= \eta,\;\\
1''&:& s_{1''}= \eta^2,\\
1'\times {1''}=1 &:&s_{1} =1
\ea

\subsection{ The cases $p=5$ }

Next we elaborate the case of $PSL_2(5)$  which is isomorphic to the symmetry group $I$ of the Dodecahedron and 
 Icosahedron as well as to $A_5$. The group $SL_2(5)$ is isomorphic to the symmetry group $2I$ of the Binary Icosahedron.
 The 60 elements of $A_5$  are generated by two generators $a,b$  with the properties
\[ a^2=b^3=(ab)^5=I\]
With the above method we find two representations of $SL_2(5)$, one of three and a second one of two dimensions.

The first generator is a unitary $3\times 3$ matrix
\ba 
A^{[3]}&=&-\frac{1}{\sqrt{5}}
\left(\begin{array}{ccc}
1&\sqrt{2}&\sqrt{2}\\	
\sqrt{2}&	\eta+\eta^4&\eta^2+\eta^3\\
\sqrt{2}&	\eta^2+\eta^3&\eta+\eta^4\\
\end{array}\right)=-\frac{1}{\sqrt{5}}
\left(
\begin{array}{ccc}
 1 & \sqrt{2} & \sqrt{2} \\
 \sqrt{2} & \frac{\sqrt{5}-1}{2}  &
   - \frac{\sqrt{5}+1}{2} \\
 \sqrt{2} &- \frac{\sqrt{5}+1}{2}  &
    \frac{\sqrt{5}-1}{2}  \\
\end{array}
\right)
\ea 
where in the last form,  the matrix elements have been written in terms of the golden ratio, since
\[    \eta+\eta^4 =\frac 12\left(\sqrt{5}-1\right),\; \eta^2+\eta^3 =-\frac 12\left(\sqrt{5}+1\right)\]
The character of the representation is Tr$A^{[3]}=-1$, as expected from the character table of $PSL_2(5)$ .

The second generator has the following three dimensional representation
\ba 
B^{[3]}&=&-\frac{1}{\sqrt{5}}
\left(
\begin{array}{ccc}
 1 & \sqrt{2} \eta ^2 & \sqrt{2} \eta ^3 \\
 \sqrt{2} & \eta ^3+\eta  & \eta +1 \\
 \sqrt{2} & \eta ^4+1 & \eta ^4+\eta ^2 \\
\end{array}
\right)
\ea 
while the character is Tr$B^{[3]}\propto 1+\eta+\eta^2+\eta^3+\eta^4=0$.
 It can be readily checked that 
$A^{[3]}$ and $B^{[3]}$ satisfy the defining relations of the $PLS_2(5)$ group:
 \[{A^{[3]}}^2={B^{[3]}}^3=({A^{[3]}\cdot B^{[3]}})^5=I\] 
These generators correspond to $3'$ triplet. Indeed, in order to make 
contact with the form of generators given in recent  literature
we transform the above in the $s_5', t_5'$ basis~\footnote{See for example neutrino models 
with $A_5$ family symmetry  in~\cite{Feruglio:2011qq,Ding:2011cm}.}, setting
\[  s_5'\equiv A^{[3]}, \;  t_5' = {A^{[3]}\cdot B^{[3]}}\; \to \;  B^{[3]}=s_5'\cdot t_5'\]
Hence, the two new generators $s_5',\, t_5'$ are
\ba 
s_5'&=&-\frac{1}{\sqrt{5}}
\left(\begin{array}{ccc}
1&\sqrt{2}&\sqrt{2}\\	
\sqrt{2}&	\eta+\eta^4&\eta^2+\eta^3\\
\sqrt{2}&	\eta^2+\eta^3&\eta+\eta^4\\
\end{array}\right),\;
t_5'=
\left(
\begin{array}{ccc}
 1 & 0 & 0 \\
 0 & n^2 & 0 \\
 0 & 0 & n^3 \\
\end{array}
\right)\label{s5t5irreps}
\ea 
They satisfy  the defining  relations
\[  {s_5'}^2 = {t_5'}^5={(s_5'\cdot t_5'})^3=I\]
while their characters are
\[ \chi_{s_5'}=-1,\; \chi_{t_5'}=\frac{1-\sqrt{5}}2 \]

\noindent
It is possible to get the other triplet representation 
of $SL_2(5)$ (up to equivalence) given in~\cite{Feruglio:2011qq}, 
\ba 
s_5&=&\frac{1}{\sqrt{5}}
\left(\begin{array}{ccc}
1&-\sqrt{2}&-\sqrt{2}\\	
-\sqrt{2}&	\eta^2+\eta^3&\eta+\eta^4\\
-\sqrt{2}&	\eta+\eta^4&\eta^2+\eta^3\\
\end{array}\right),\;
t_5=
\left(
\begin{array}{ccc}
 1 & 0 & 0 \\
 0 & n^2 & 0 \\
 0 & 0 & n^3 \\
\end{array}
\right)
\ea 
by redefining $\eta$ to $\eta^3$ in eq.(\ref{s5t5irreps}).
As discussed in section 4  after eq.(\ref{UAelms}) this is equivalent to a
rescaling of  appropriate elements of $SL_2(5)$ which give the generators $s_5'$ and $t_5'$.

\noindent
The two-dimensional representation gives
\ba 
A^{[2]}&=&-\frac{1}{\sqrt{5}}
\left(\begin{array}{cc}	
	\eta^4-\eta&\eta^2-\eta^3\\
	\eta^2-\eta^3&\eta-\eta^4\\
\end{array}\right)
\ea 
and
\ba 
B^{[2]}&=&-\frac{1}{\sqrt{5}}
\left(\begin{array}{cc}	
	\eta^2-\eta^4&\eta^4-1\\
	1-\eta&\eta^3-\eta\\
\end{array}\right)
\ea 
satisfying 
 \[{A^{[2]}}^2={B^{[2]}}^3=({A^{[2]}\cdot B^{[2]}})^5=-I\] 
Obviously, the above two-dimensional matrices are  representations of $SL_2(5)$, but  not of 
$PSL_2(5)$.

The three and the two-dimensional representations of the generators constructed above,
are unitary matrices and so they generate discrete subgroups of $SU(3)$ and $SU(2)$ Lie groups.

\subsection{ The cases $p=7$ }

As a final example in this note, we consider the case $p=7$. The associated 
groups are $SL_2(7)$  with  336 elements and its projective one
 $PSL_2(7)$ which has 168 elements and it is a discrete simple subgroup  of $SU(3)$.
It is the group preserving the discrete  projective geometry of the Fano plane realising
 the multiplication structure of the octonionic units.

 \noindent
Using the method described above we will construct the four- and three-dimensional 
representations of  $SL_2(7)$. The three-dimensional one is also a  representation of $PSL_2(7)$
\[   a^2=b^3 =(ab)^7= [a,b]^4=1  \]
with $[a,b]=a^{-1}b^{-1}ab$.

The $A^{[4]}$ and $B^{[4]}$ generating matrices of  the irreducible four-dimensional unitary representation
of $SL_2(7)$ are 
\ba 
A^{[4]}&=&\frac{i}{\sqrt{7}}
\left(\begin{array}{cccc}
	1&	\sqrt{2}&\sqrt{2}&\sqrt{2}\\
\sqrt{2}&	\eta+\eta^6&\eta^2+\eta^5&\eta^3+\eta^4\\
\sqrt{2}&	\eta^2+\eta^5&\eta^3+\eta^4&\eta+\eta^6\\
\sqrt{2}&	\eta^3+\eta^4&\eta+\eta^6&\eta^2+\eta^5
\end{array}\right)
\ea 
and
\ba 
B^{[4]}&=&\frac{i}{\sqrt{7}}
\left(\begin{array}{cccc}
	1&	\sqrt{2}\eta^3&\sqrt{2}\eta^5&\sqrt{2}\eta^6\\
	\sqrt{2}&	\eta^2+\eta^4&1+\eta^3&\eta^2+\eta^3\\
	\sqrt{2}&	\eta+\eta^5&\eta+\eta^2&1+\eta^5\\
	\sqrt{2}&	1+\eta^6&\eta^4+\eta^6&\eta+\eta^4
\end{array}\right)
\ea

These matrices satisfy the relations for $SL_2(7)$ which are
\[{A^{[4]}}^2={B^{[4]}}^3=({A^{[4]}B^{[4]}})^7=[{A^{[4]},B^{[4]}}]^4=-I\]

The generators in the triplet representation are the following
\ba 
A^{[3]}&=&\frac{i}{\sqrt{7}}
 \left(\begin{array}{ccc}
\eta^2-\eta^5&\eta^6-\eta&\eta^3-\eta^4\\
\eta^6-\eta&\eta^4-\eta^3&\eta^2-\eta^5\\
\eta^3-\eta^4&\eta^2-\eta^5&\eta-\eta^6
\end{array}\right)
\ea 
and
\ba 
B^{[3]}&=&\frac{i}{\sqrt{7}}
\left(\begin{array}{ccc}
	\eta-\eta^4&\eta^4-\eta^6&\eta^6-1\\
	\eta^5-1&\eta^2-\eta&\eta^5-\eta\\
	\eta^2-\eta^3&1-\eta^3&\eta^4-\eta^2
\end{array}\right)
\ea 

As expected, the $A^{[3]}$ and $B^{[3]}$  satisfy the defining relations

\be 
{A^{[3]}}^2={B^{[3]}}^3=({A^{[3]}B^{[3]}})^7=[{A^{[3]},B^{[3]}}]^4=I
\ee 
We note that our  representations $A^{[3]}, B^{[3]}$ are connected to the conjugate triplet 
of those of refs~\cite{Luhn:2007sy,King:2009mk} through the similarity transformation obtained by 
 the diagonal matrix ${\tiny (1,-1,-1)}$.
We also note in passing that the phenomenological implications of $PSL_2(7)$ have been analysed in several 
works~(see reviews~\cite{Altarelli:2010gt,Ishimori:2010au,King:2013eh} and references therein).


\section{Conclusions}

In the present note, we have introduced an intriguing relation of the discrete flavour symmetries with the 
 automorphisms   of the magnetic translations of the finite and discrete Heisenberg Group.
This relation is reminiscent of the discrete symmetries of the Quantum Hall effect, where in a toroidal two dimensional 
space  the magnetic flux transforms the torus to a phase space and the Hilbert space of a charged particle becomes 
finite dimensional and the corresponding torus effectively discrete~\cite{Zak:1989jz}. 
Torii with fluxes in internal extra  dimensions appear naturally in the framework 
of F-theory of elliptic fibrations over Calabi-Yau manifolds, where they generate the GUT gauge groups
 and  other discrete symmetries at particular singularities of the fibration.  Phenomenological 
explorations have shown that such discrete symmetries are particularly successful 
in predicting the fermion mass hierarchies and the flavour mixing.
Inspired by these observations we made use of the discrete Heisenberg Group to  develop
 a simple and unified  method for the derivation  of  basic non-trivial representations of a large class of 
 non-abelian finite groups relevant to the flavour symmetries.  It will be important to construct explicit 
models of elliptic fibrations with fluxes, where the discrete magnetic translations appear naturally and the
 discrete flavour symmetries as their automorphisms.

\vfill 

{\it The authors would like to thank CERN theory division for kind hospitality where
 this work  has been done. We would like to express our deep sorrow for the loss of our colleague Guido Altarelli 
who passed away recently. His work  on the area of the neutrino physics  
for  mass and mixing matrices,among many others worldwise recognized- has been very important and influential and it is exactly the area of the recent 
Nobel Prize in Physics 2015. We will miss him.}

\newpage

\end{document}